\documentclass[5p]{elsarticle}
\usepackage{amsmath}
\usepackage{graphicx}
\usepackage{epstopdf}
\usepackage{amssymb}
\usepackage{wrapfig}
\usepackage{lineno}
\usepackage{multicol}
\usepackage{booktabs}
\usepackage{url}
\usepackage{hyperref}
\usepackage{natbib}
\usepackage{notoccite}
\numberwithin{equation}{section}
\pdfpagewidth=\paperwidth
\pdfpageheight=\paperheight
\usepackage[font=scriptsize]{caption}

\makeatletter

\long\def\@makecaption#1#2{
  \vskip\abovecaptionskip
  \sbox\@tempboxa{{\captionfonts #1: #2}}%
  \ifdim \wd\@tempboxa >\hsize
    {\captionfonts #1: #2\par}
  \else
    \hbox to\hsize{\hfil\box\@tempboxa\hfil}%
  \fi
  \vskip\belowcaptionskip}
\makeatother
\newcommand{\captionfonts}{\scriptsize}

\begin{document}

\begin{frontmatter}
\title{Cross Calibration of Telescope Optical Throughput Efficiencies using Reconstructed Shower Energies for the Cherenkov Telescope Array}

\author[37]{A.~M.~W.~Mitchell\corref{cor1}} 
\ead{Alison.Mitchell@mpi-hd.mpg.de}
\author[37]{R.~D.~Parsons} 
\author[37]{W.~Hofmann} 
\author[37]{K.~Bernl\"{o}hr} 
\cortext[cor1]{Corresponding Author}

\address[37]{Max-Planck-Institut f\"{u}r Kernphysik, Germany}

\begin{abstract}
For reliable event reconstruction of Imaging Atmospheric Cherenkov Telescopes (IACTs), calibration of the optical throughput efficiency is required. Within current facilities, this is achieved through the use of ring shaped images generated by muons. Here, a complementary approach is explored, achieving cross calibration of elements of IACT arrays through pairwise comparisons between telescopes, focussing on its applicability to the upcoming Cherenkov Telescope Array (CTA). 
Intercalibration of telescopes of a particular type using eventwise comparisons of shower image amplitudes has previously been demonstrated to recover the relative telescope optical responses. A method utilising the reconstructed energy as an alternative to image amplitude is presented, enabling cross calibration between telescopes of varying types within an IACT array. Monte Carlo studies for two plausible CTA layouts have shown that this calibration procedure recovers the relative telescope response efficiencies at the few percent level.

\end{abstract}


\end{frontmatter}

\section{Introduction}
\label{sec:intro}
\noindent Very High Energy (VHE, $\gtrsim O$(50~GeV)) $\gamma$-ray astronomy is concerned with energetic photon signals from some of the highest energy astrophysical processes, such as blazar jets, pulsar wind nebulae and supernova remnants \citep{HintonHofmann09}. Ground-based experiments observe these astrophysical VHE $\gamma$-ray photons through extensive air showers (EAS) generated by their interaction with the Earth's atmosphere. Charged particles within these air showers travelling faster than the local speed of light in air produce Cherenkov radiation, detectable at ground level. Current experiments such as HESS \cite{Hinton04}, MAGIC \cite{Lorenz04} and VERITAS \cite{Weekes02} use arrays of Imaging Atmospheric Cherenkov Telescopes (IACTs) to record images of this Cherenkov radiation, from which the primary particle type, energy and direction of origin may be reconstructed, with $\gamma$-rays generating characteristically elliptical images.

\noindent As energy reconstruction is directly related to the intensity of the shower image recorded, a reliable calibration of telescope optical throughput efficiency, relating a number of recorded digital counts to the original light yield, is required for source flux and energy interpretation. Telescope optical throughput efficiency (or equivalently, the optical response efficiency as a percentage of the nominal value) is measured as a cumulative effect of several contributing factors, including: mirror reflectivity, shadowing of the mirror due to the telescope structure, photoelectron charge collection efficiency and quantum efficiency of the photodetectors. Knowledge of this efficiency parameter is required for the shower reconstruction and to account for degradation of telescope components over time. 

\noindent This parameter can be determined via ring--like images generated by background muons. The amount of light emitted by a muon is analytically calculable from geometrical parameters of the ring image formed \cite{Vacanti94,Bolz04}. Fitting the ring--like image to determine the geometrical parameters (ring radius, ring width, location of muon impact) and comparing the image charge recorded in photoelectrons to the theoretical expectation for the same ring geometry enables the telescope--wise optical throughput efficiency to be determined. This procedure is performed within the current generation of IACTs on an individual telescope basis, under the assumption that the measured Cherenkov light spectrum from $\gamma$-ray showers and muons is the same. In practice, upon arrival at telescope level, the Cherenkov light spectrum due to muons is bluer than that due to $\gamma$-rays as a consequence of the lower production altitude.

\noindent The next generation Cherenkov Telescope Array (CTA) facility will comprise at least three different telescope sizes and designs of telescope \cite{Acharya13}, potentially incorporating both traditional single reflector, such as Davies--Cotton (DC) \cite{DaviesCotton57} and parabolic, and novel dual mirror Schwarzschild--Couder (SC) \cite{Vassiliev08} optical designs. This will enable enhanced background rejection, angular resolution and energy resolution over current facilities. To achieve these improvements, robust calibration of telescope relative response efficiencies globally across a whole array is important. Although the use of a muon efficiency calibration is envisaged on all telescopes individually within CTA, in multi-size arrays, there can be wide variation in trigger rates; of particular concern is a potentially low muon trigger rate expected for the small-size telescopes (SSTs), with primary mirror diameters of around 4~m.  In comparison, the expected mirror diameters for the medium (MSTs) and large (LSTs) telescopes are of order 12~m and 23~m respectively, such that no issues arising from a low muon trigger rate are foreseen.

\noindent We present here a method of telescope cross calibration (between telescopes of different types), building on the intercalibration (between telescopes of the same type) approach first outlined in \cite{Hofmann03}, enabling the optical throughput efficiency of individual telescopes to be measured. In this previous study on the HEGRA array, photon shower events well above threshold falling approximately equidistant between two telescopes were selected as those which are expected to generate identical responses. The recorded image sizes (the total photoelectron amplitude of all pixels contained within a cleaned image) were then compared pairwise between telescopes, with the responses of the telescopes being intercalibrated at a level of 1--2\%. Similar levels of accuracy have also been demonstrated in applications of the method to the H.E.S.S. and MAGIC arrays \cite{Gast09,Aleksic12}. Therefore, this approach has the potential to calibrate the telescope optical efficiencies at the few percent level, fulfilling the CTA performance goal of $5\%$ systematic uncertainty on the absolute intensity of Cherenkov light at each telescope \cite{Gaug14}. 

\noindent Whilst comparison of image sizes directly is possible for telescopes of identical specifications, in the case of CTA with multiple telescope types, the variation in image size with hardware cannot be corrected for, especially near trigger threshold. Although use of an expected size ratio (obtained from simulations) may be considered, this approach suffers considerably from low statistics and random fluctuations for telescope separations exceeding the radius of the Cherenkov light pool $\sim O(100~\mathrm{m})$.

\noindent Instead, we propose an approach using the reconstructed shower energy as an alternative to the image size, as beyond the biased threshold region all telescopes should agree on the shower energy. The cross calibration is performed in a relative manner, with a standard Hillas based analysis procedure used for the energy reconstruction \cite{Hillas85}, although this calibration could also be performed within a template based analysis framework \cite{Lebohec98}. Building a system of measurements by telescopes participating in multiple pairwise comparisons enables an overdetermination of unknown parameters, and hence an entire array to be precisely calibrated. This procedure also circumvents any systematic bias arising from the difference in Cherenkov light spectrum produced by muons and $\gamma$-ray showers by calibrating using the $\gamma$-ray showers themselves.

\section{Cross Calibration Principle}
\label{sec:CrossCalibMethod}

\noindent Shower events triggering multiple telescopes are reconstructed by each telescope pair separately. Using additional telescopes which triggered on the same event in the shower reconstruction can introduce a bias in the calibration which could skew the results obtained. The variation in reconstructed energy between two telescopes is quantified by the use of an asymmetry parameter, $a_{ij}$:
\begin{equation}
a_{ij} = \frac{E_i - E_j}{E_i + E_j}~,
\label{eq:asymmetry}
\end{equation}
where $E_i$ and $E_j$ are the energies reconstructed by telescopes $i$ and $j$ individually. From the image size and reconstructed distance, a per telescope energy estimate and uncertainty is obtained from a set of lookup tables. 
\noindent In the case of two telescopes of the same type, the energy asymmetry distribution is symmetric; however, when comparing telescopes of different types a biased region occurs near the trigger threshold of the smaller telescope. Hence measurements of the energy asymmetry are only made between telescopes of the same type.

\noindent The overall asymmetry $a_{ij}$ and uncertainty $\sigma_{ij}$ between two telescopes is determined by the weighted mean and uncertainty of all available event-wise energy asymmetry estimates, weighted by $\sigma_{ij}^{-2}$. This energy asymmetry is interpreted as corresponding to the intrinsic optical efficiency asymmetry of the two telescopes, with all other factors not mentioned contributing at a negligible level.

\begin{figure}
\begin{center}
	\includegraphics[width = \columnwidth]{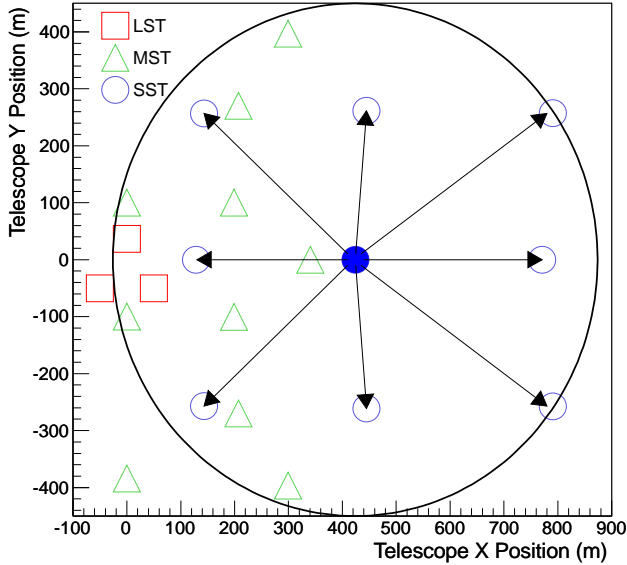}
	\caption{Part of a CTA array layout. For a given telescope (filled in circle), asymmetry measurements are made with all other telescopes of the same type falling within a certain radius.}
	\label{fig:CTAMinSchematic}
\end{center}
\end{figure}

\noindent Measurements for the optical response asymmetry between two telescopes are made for all pairs of telescopes of the same type. In the case of large arrays, a limit on the separation distance of telescopes between which measurements are made needs to be imposed. Once the telescope separation distance exceeds the typical Cherenkov light pool diameter, the reconstruction performance drops off with increasing distance; the much lower rate of events triggering both telescopes leads to lower statistics, such that comparisons between telescopes at large separation distances are impractical. Nevertheless, with a cut on separation distance the majority of telescopes within an array still participate in multiple asymmetry measurements $a_{ij}$, such that the system of unknown parameters (response coefficient $c_i$ per telescope $i$) is overdetermined. These response coefficients can therefore be recovered via a $\chi^2$ minimisation procedure for all valid pairs up to $N$ telescopes:

\begin{equation}
\chi^2 = \sum_{i=1, j>i}^N \frac{\left(a_{ij}-\frac{c_i-c_j}{c_i+c_j}\right)^2}{\sigma_{ij}^2}~.
\label{eq:chisquare}
\end{equation}
This minimisation procedure is performed on each subsystem of telescopes of a given type separately. An example system of measurements for a given telescope is depicted schematically in figure \ref{fig:CTAMinSchematic}.

\noindent During the minimisation, one telescope is arbitrarily kept fixed in order to remove free parameters due to the overall system scaling factor. As all measurements are made in a relative sense between telescope pairs,  the telescope-wise optical response efficiencies obtained for each telescope subsystem may be normalised after minimisation in a second step outlined below.

\section{Monte Carlo Studies}
\label{sec:MCstudies}
\begin{figure*}
	\includegraphics[width=\columnwidth]{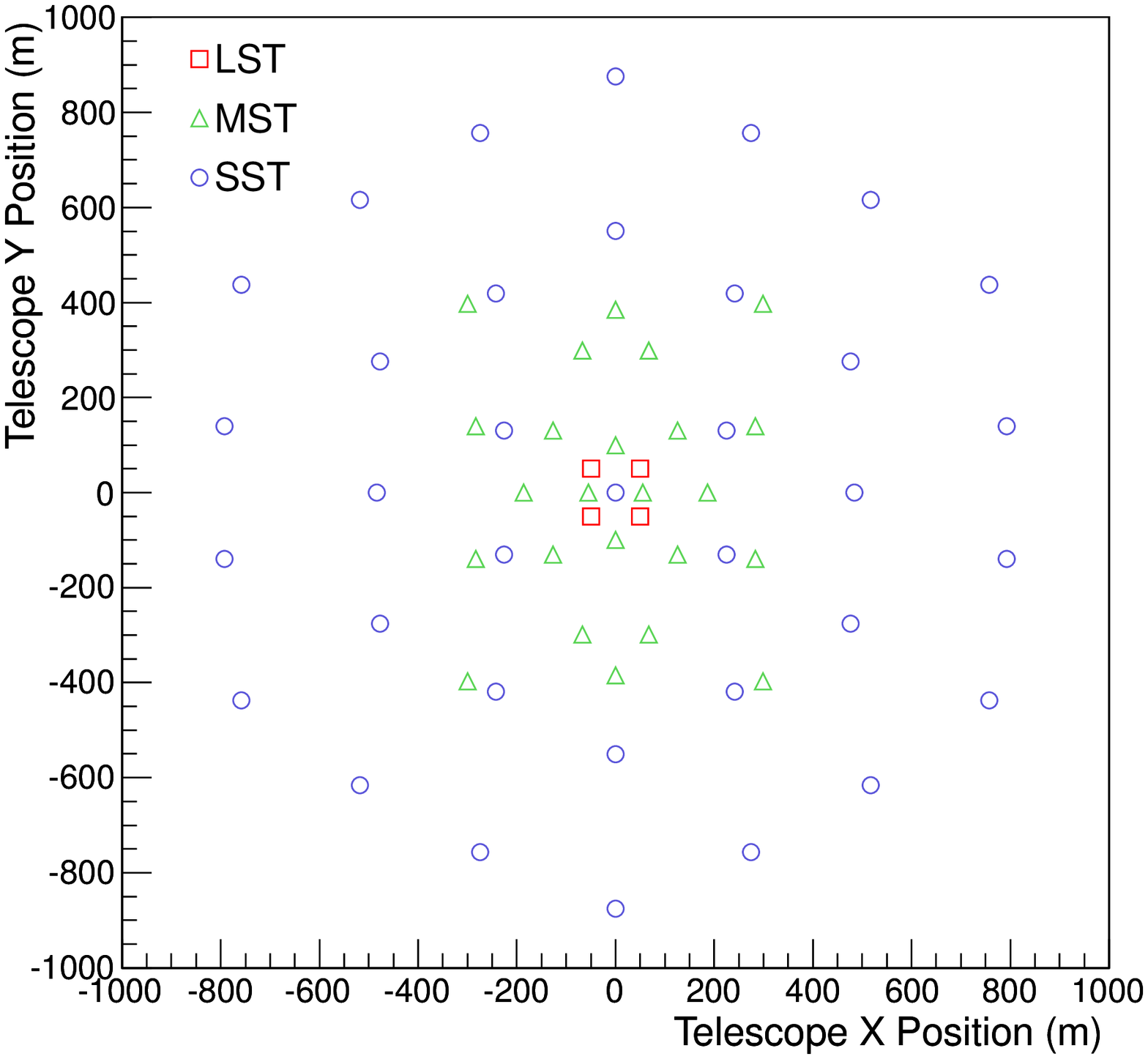}
	\includegraphics[width=\columnwidth]{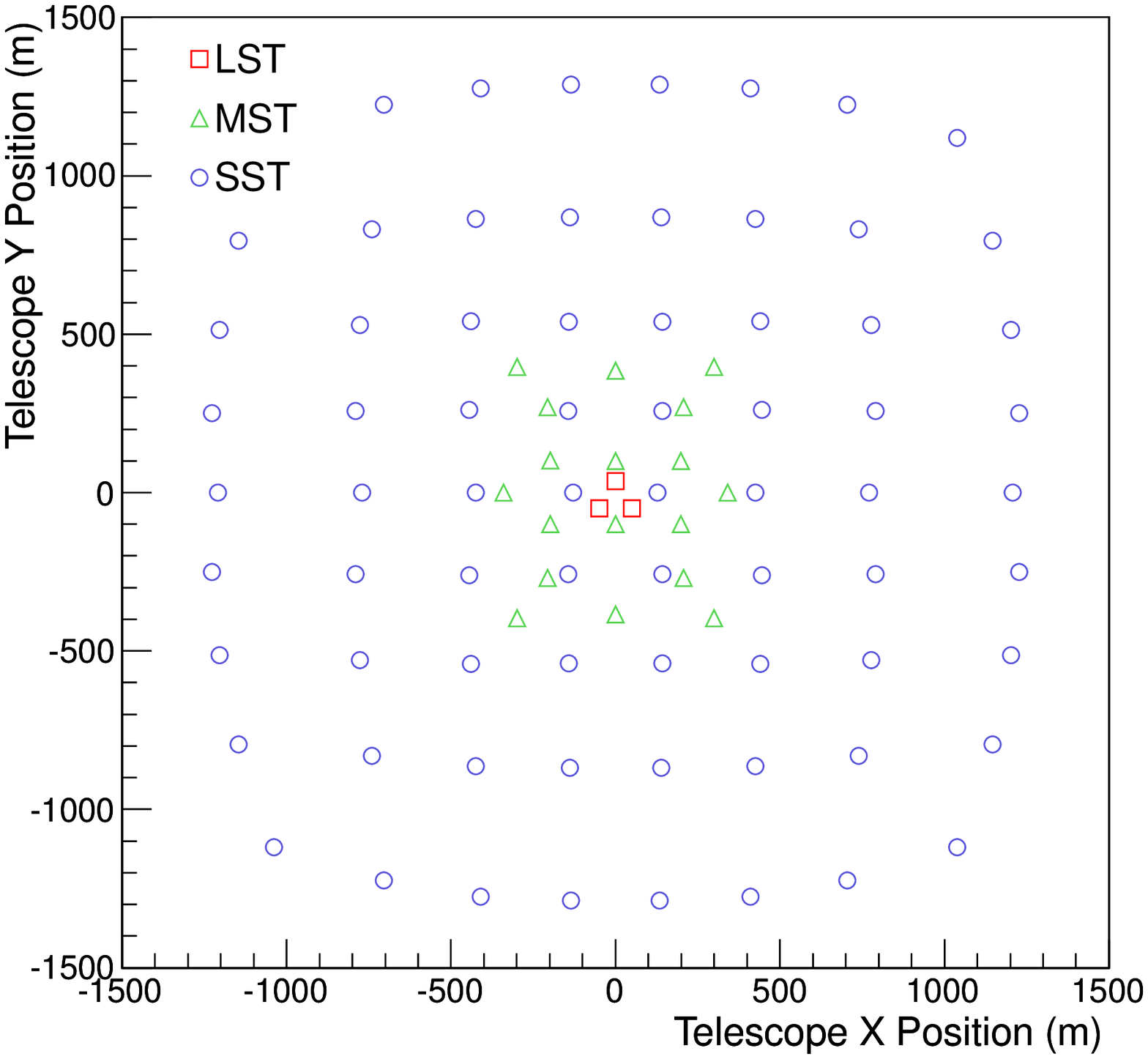}
	\caption{Array layouts used in the MC studies performed, with number of telescopes and primary mirror diameters as follows. Left: Layout 2A comprising 63 telescopes; 4 LSTs (23~m), 24 MSTs (12~m) and 35 DC-SSTs (7~m). Right: Layout 2B comprising 93 telescopes; 3 LSTs (23~m), 18 MSTs (12~m) and 72 SC-SSTs (4~m).}
	\label{fig:layouts}
\end{figure*}

\noindent As a test of the method outlined above, Monte Carlo (MC) simulations were performed using the second CTA production configuration of CORSIKA/\textit{sim$\_$telarray} \cite{Bernlohr13ICRC}. One of the candidate Southern sites for CTA was simulated with $\gamma$-rays at 20$^{\circ}$ zenith angle. From these simulations, two potential array layouts, 2A and 2B were selected as depicted in figure \ref{fig:layouts}. Both layouts have similar LST and MST configurations, whilst the SST design and configuration differs. In the case of layout 2A, the SSTs have 7~m diameter mirrors and Davies Cotton optics \cite{DaviesCotton57} (DC-SST), whereas the SSTs in the 2B array layout are of Schwarzschild-Couder optical design \cite{Vassiliev08} with a 4~m diameter primary mirror (SC-SST) and cover a wider ground area in order to maximise the effective area at the highest energies. However, it should be noted that whilst development of the DC-SST for CTA has progressed on to telescopes of 4~m diameter rather than the 7~m diameter of layout 2A, this does not affect our conclusions on the applicability of this method for either intercalibration or cross calibration. 

\noindent All telescopes were randomly assigned optical response efficiency values according to a Gaussian distribution with a mean response efficiency of 70$\%$ nominal value and a standard deviation of 10$\%$, limited to values $\leq 100\%$.  Gamma ray initiated air showers originating from 20$^{\circ}$ zenith angle were subsequently simulated according to an $E^{-2}$ energy spectrum over an energy range of $\sim 5~\mathrm{GeV}-500~\mathrm{TeV}$. 

\noindent These simulated events were reconstructed separately by all independent same type telescope pairs as outlined above using the CTA baseline analysis \citep{Bernlohr13}. The reconstruction was performed using a set of lookup tables based on all telescopes having 70$\%$ optical efficiency. 
 
\noindent Measurements were made between all telescope pairs with separation distances $d_{ij}< 300~\mathrm{m}$, enabling many cross checks (table \ref{tab:ntelmeas}). For the SSTs, which are spread over a larger area (figure \ref{fig:layouts}), the allowable telescope separation was increased to $d_{ij}<350~\mathrm{m}$ for the 7~m DC-SSTs of array 2A, and to $d_{ij}<450~\mathrm{m}$ for the 4~m SC-SSTs of array 2B, permitting cross checks in the outermost regions of the array.

\begin{table}
\begin{center}
\begin{tabular}{ c c c c }
\toprule
Array & Subsystem & Telescopes & Telescope Pairs\\
\midrule
 & LST & 4 & 6 \\
2A & MST & 24 & 100 \\
 & DC-SST & 35 & 78 \\
\midrule
 & LST & 3 & 3 \\
2B & MST & 18 & 37 \\
 & SC-SST & 72 & 183 \\
\bottomrule
\end{tabular}
\end{center}
\caption{Number of telescopes and of telescope pairs suitable for asymmetry measurements for each telescope type subsystem. }
\label{tab:ntelmeas}
\end{table}

\noindent To ensure that the shower events used for the calibration are treated in a consistent manner, are sufficiently above threshold to avoid instrumental effects and measurement biases are minimal, a set of good quality cuts were defined as outlined in table \ref{tab:cuts}.
 
\begin{table}
\begin{center}
\begin{tabular}{ l c c c c }
\toprule
Parameter & LST & MST & DC-SST & SC-SST \\
\midrule
Image Size (pe) & 93 & 90 & 79 & 29\\
Tail Cuts (pe) & 9,12 & 8,11 & 6,9 & 3.7,5.5 \\ 
Pixels in image & 5 & 4 & 3 & 4 \\
\bottomrule
\end{tabular}
\end{center}
\caption{Event selection cuts used for the cross-calibration 
procedure. All values quoted are minimum thresholds (see \citep{Bernlohr13}), determined for the central energy range according to telescope dependent properties.The tail cuts image cleaning retains all pixels with photoelecton charge greater than the upper threshold and neighbouring pixels with charge greater than the lower threshold.}
\label{tab:cuts}
\vspace{-2mm}
\end{table}

\noindent  Additionally, weak background cuts on the image shape were imposed; only events with a mean reduced scaled width and mean reduced scaled length (scale parameters characterising images from multiple telescopes in stereoscopic observations) of less than 2 were included \citep{Bernlohr13}. All shower events were required to trigger at least two telescopes.

\section{Simulation Results}
\label{sec:simresults}
\subsection{Asymmetry Correlation}
\noindent Independent asymmetry measurements were made between telescope pairs per subsystem as summarised in table \ref{tab:ntelmeas}. 

\begin{figure*}
\includegraphics[width = \columnwidth]{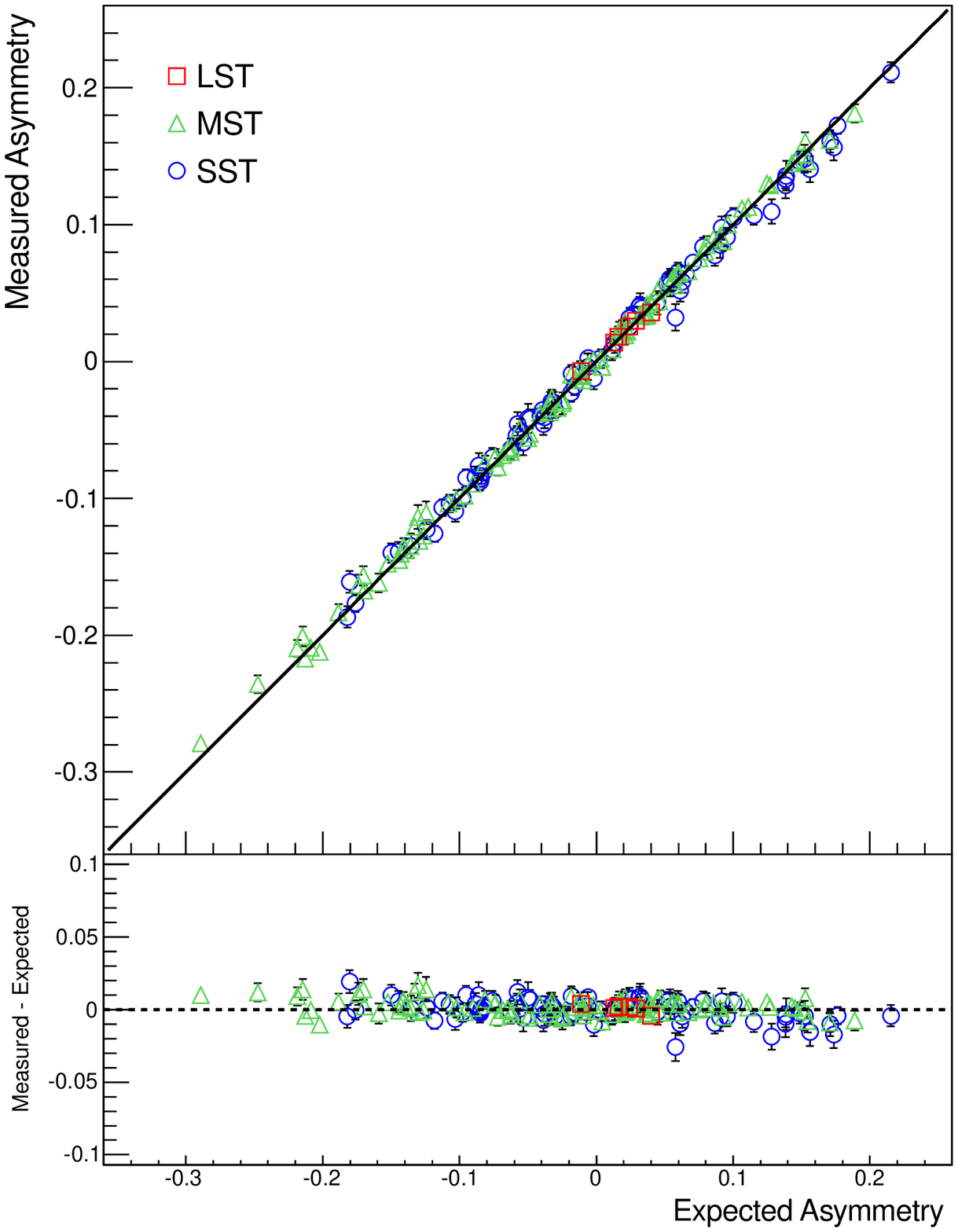}
\includegraphics[width = \columnwidth]{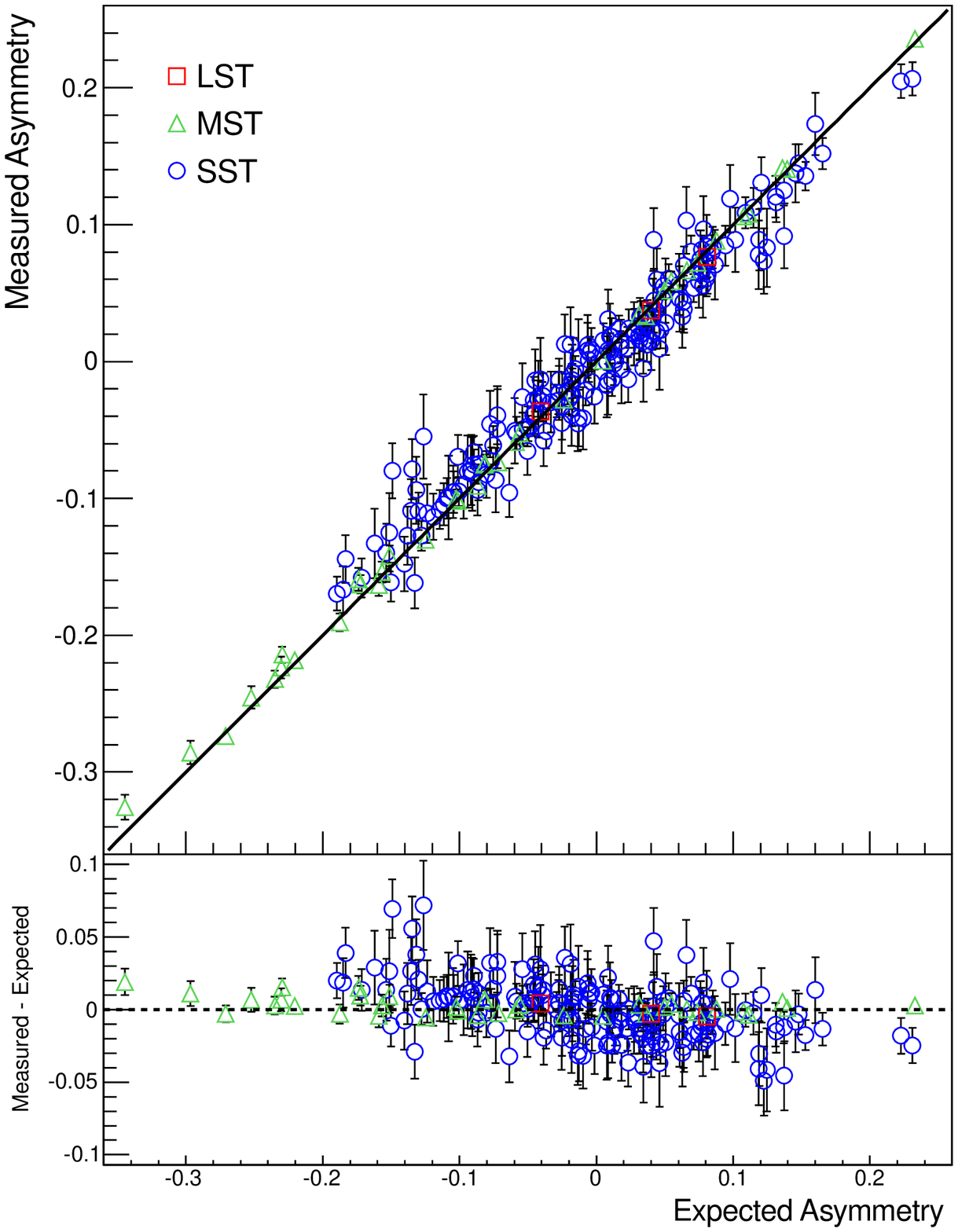}
\caption{Measured asymmetries against the expected values calculated from the preassigned optical efficiencies, for the array layouts 2A (left) and 2B (right). Shown for all telescope pairs with pairs between LSTs shown in red, between MSTs in green and between SSTs in blue. Measurement errors are typically of the order of the point size $\sim O(0.005)$, noticeably increased (to $\sim O(0.01)$) for SST pairs in the outermost regions of layout 2B. A black, solid line defines the case of perfect agreeement.}
\label{fig:asymagreement}
\end{figure*}

\noindent Excellent agreement between the obtained reconstructed energy asymmetry and the true expected response efficiency asymmetry (the asymmetry of the preassigned telescope mirror reflectivities) is demonstrated in figure \ref{fig:asymagreement}, verifying that the reconstructed energy is a valid alternative variable to the image size and does indeed probe the telescope response asymmetry. Larger errors are consistently obtained for the smaller 4~m SSTs in layout 2B than for the 7~m SSTs in layout 2A, especially for telescope pairs at larger separations in the outermost regions of the array, where the event rate (and hence statistics) is particularly low. Correspondingly, there are fewer events triggering both telescopes due to the increased telescope separation distance, causing lower statistics and larger measurement errors.

\noindent At large asymmetries some small bias was found, mainly within the MST subsystem, as seen particularly for layout 2B in figure \ref{fig:asymagreement}; on further investigation, these pairs involved one of two MSTs located centrally to the array layout with randomly assigned mirror reflectivities of 49\% and 100\% of nominal, quite far from the mean of 70\% used in the lookup tables. Consequently, energy asymmetry distributions produced with these telescopes were slightly biased due to the $20-30\%$ discrepancies in optical throughput efficiency involved. This bias would be reduced in the case of an iterative approach. Nevertheless, the overall correlation between the true optical efficiency asymmetry and the measured energy asymmetry is excellent.

\subsection{Subsystem Minimisation}
\label{sec:subsysmin}

\noindent The individual telescope response efficiencies were found using the aforementioned procedure, assuming that all telescopes had a response efficiency of 70$\%$ as starting points for the minimisation of each subsystem. For the minimisation, performed using MINUIT \cite{James75}, one telescope was arbitrarily kept fixed, to avoid ambiguities due to the uncertainty in the overall scaling factor of the system. This restriction is later lifted, as outlined below.

\noindent  In order to cross calibrate the entire array, the relative normalisation between the telescope type subsystems is required. As the minimisation of each subsystem was performed in a relative manner, the average response efficiency per subsystem was defined to be equal to 1.0 and the telescope--wise relative response efficiencies were correspondingly rescaled. The reciprocal of these rescaled efficiencies gave a per telescope correction factor, which was applied to the per telescope energy estimates in order to correct for the differences in efficiencies as found from the minimisation. Subsequently, the relative normalisation between the telescope subsystems were obtained as follows.

\subsection{Scaling Factors between Subsystems}
\label{sec:scalingfactors}

\noindent Events detected by two different telescope types, with at least two telescopes of each type triggered, were selected. These showers were reconstructed separately by each telescope type subsystem, with all telescope energy estimates corrected for efficiency variation within each subsystem as outlined above. The mean ratio of the event--wise subsystem energy estimates, obtained by a Gaussian fit to the distribution accumulated from the same data set, provided the scaling factors between subsystems. 

\noindent As the two different array layouts comprise different combinations of telescopes with differently assigned efficiencies, as well as different SST types, the average true efficiency of the various telescope type subsystems does not remain the same. The scaling factors adjust for this, and are indeed seen to differ for the same scaling (such as LST/MST) between the two array layouts.

\noindent Due to the requirement of multiple telescopes of each type triggering in all events used to determine the scaling factors, there remains a systematic bias towards events near the centre of the array. Consequently, the scaling factors found from the Gaussian fit more closely match the mean efficiencies of centrally located telescopes than those at larger distances. To account for this, the scaling factors are modified by the ratio of the average telescope relative response efficiency to the average weighted by telescope participation. 

\noindent Using these scaling factors to scale the response efficiencies obtained for the LST and SST subsystems to the MST subsystem yielded the overall relative response efficiencies of all telescopes within an array.

\subsection{Recovered Efficiencies}
\label{sec:recoeffs}

\begin{figure*}
\includegraphics[width = \columnwidth]{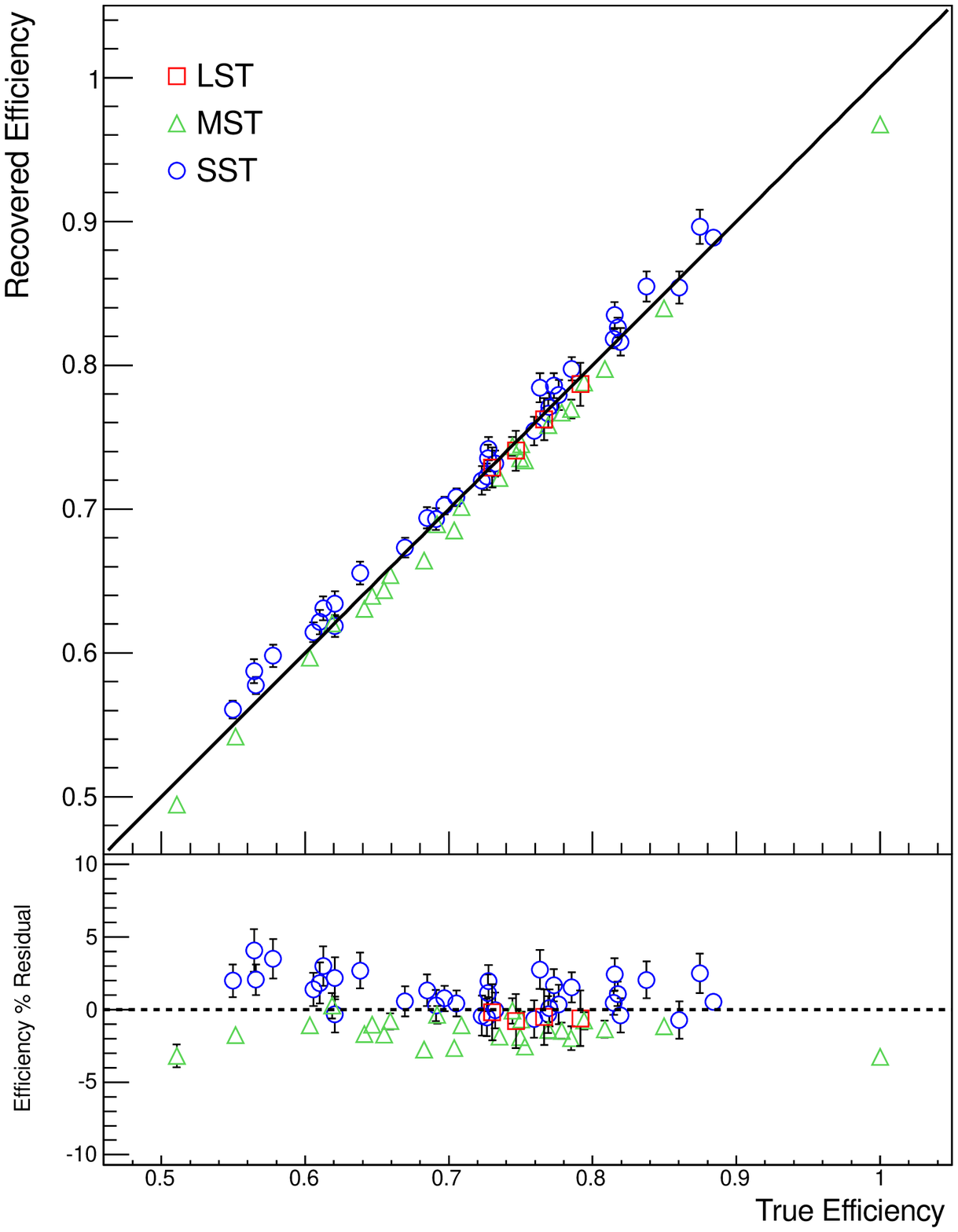}
\includegraphics[width = \columnwidth]{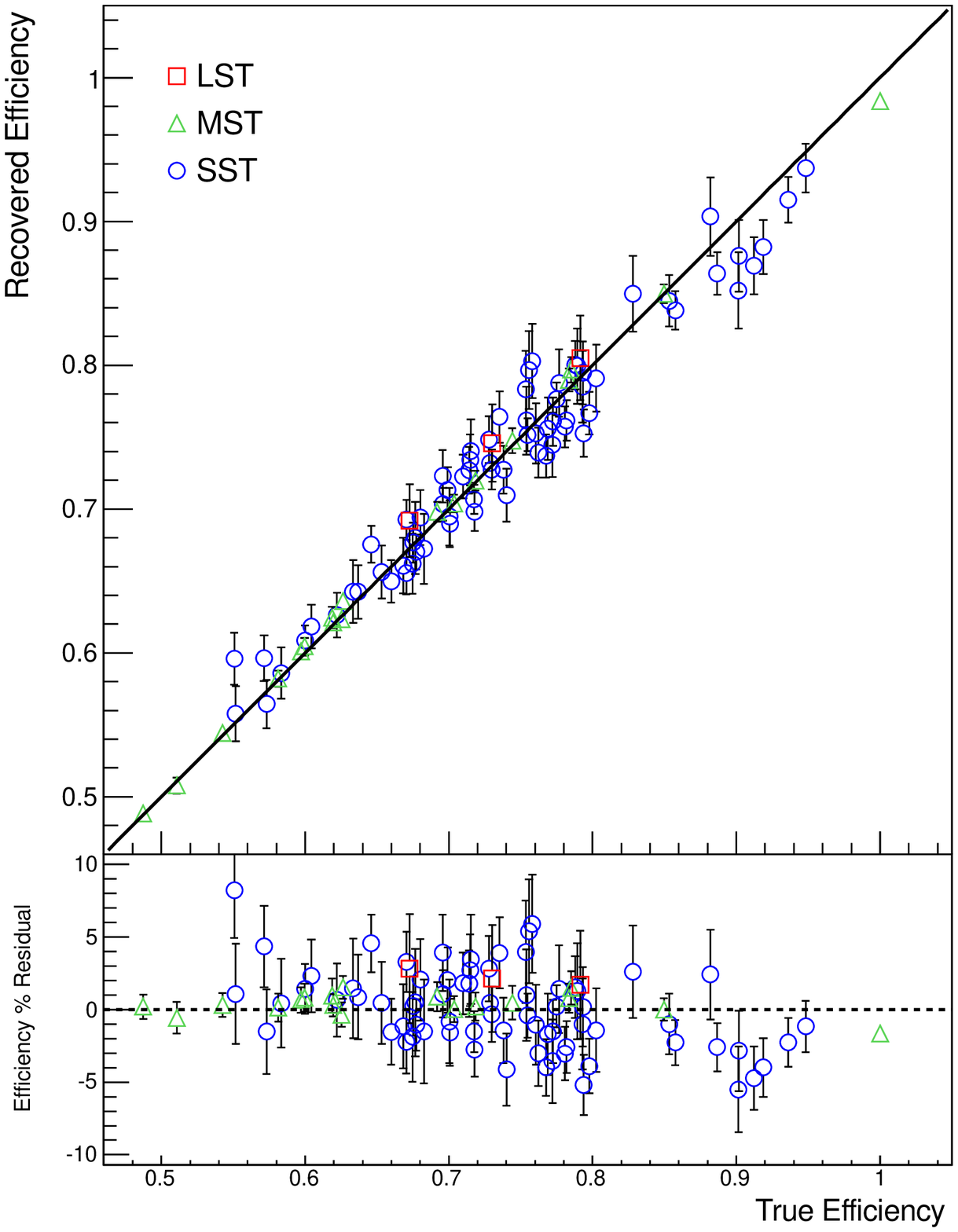}
\caption{Telescope response efficiencies obtained post minimisation compared to the initially assigned values. Results correspond to the 2A array layout (left) and the 2B array layout (right). Measurement errors are typically  $\lesssim O(0.02)$ and are statistical only, with percentage residuals shown below. A black, solid line defines the case of perfect agreeement.}
\label{fig:effagree}
\end{figure*}

\noindent The agreement obtained between the assigned and recovered telescope response efficiencies with this cross calibration procedure (after relative normalisation of telescope subsystems) is shown in figure \ref{fig:effagree}, with the RMS of the residuals for different array layouts and telescope subsystems stated in table \ref{tab:arrayrmss}. For the purposes of illustration, the mean recovered efficiency was set to the mean true preassigned efficiency; in practice, as long as the calibration procedures and simulations are self-consistent, the absolute value of this normalisation is not necessary for telescope cross calibration. 

\noindent In obtaining this level of precision, $35\times 10^{6}$ $\gamma$--ray events were simulated, with $O(1\times 10^{5})$ events triggering the  array and $O(4\times 10^{4})$ events passing cuts for both array layouts. This corresponds to approximately 10 or 13 hours of data on a source with 10\% of the $\gamma$--ray flux of the Crab Nebula (over the energy range 50~GeV --100~TeV), adopting a spectrum as measured by the H.E.S.S. experiment \cite{Aharonian06}, for array layouts 2A and 2B respectively. Accordingly, the sparser 2B array layout optimised for the rarer events at the high end of the energy range, requires longer to collect similar suitable calibration event statistics to the more compact 2A array layout.

\begin{table}
\begin{center}
\begin{tabular}{ c c c }
\toprule
Array Component & 2A & 2B \\
\midrule
 LSTs & 0.2$\%$ & 0.5$\%$ \\
 MSTs & 0.9$\%$ & 0.7$\%$ \\
 SSTs & 1.2$\%$ & 2.8$\%$ \\
 Full array & 1.7$\%$ & 2.5$\%$ \\
\bottomrule
\end{tabular}
\end{center}
\caption{Root Mean Square percentage residual for different arrays and components after simulations corresponding to approximately 10 hours (2A) or 13 hours (2B) of data on a source with 10\% of the $\gamma$--ray flux of the Crab Nebula (over the energy range 50~GeV -- 100~TeV), adopting a spectrum as measured by the H.E.S.S. experiment \cite{Aharonian06}. These percentage residuals are calculated for each subsystem independently. The full array value corresponds to figure \ref{fig:effagree}, after all subsystem scaling factors have been applied.}
\label{tab:arrayrmss}
\end{table}

\noindent Although there was a small bias seen at large asymmetries for some MST pairs, this did not adversely affect the resulting overall calibration, as demonstrated by the excellent agreement shown in figure \ref{fig:effagree}. Some bias is seen for those telescopes with particularly high or low efficiencies; this is due to reconstructing the event energy from lookup tables with all telescope efficiencies set at 70\% of nominal value, quite far from the true value in these extreme cases. In practise, this calibration would be performed in an iterative manner, with lookup tables generated based either on telescope efficiencies obtained with a first pass attempt, calibration from the previous month or from a muon calibration, thereby reducing the influence of these biases.

\noindent Additionally, a small systematic bias in the scaling between telescope subsystems is also seen in figure \ref{fig:effagree}, despite adjusting the scaling factors for being more heavily weighted towards telescopes in more densely packed regions of the array.

\noindent  In the case of the 2B array layout, larger scatter was seen in the SSTs than for 2A, due partly to the reduced mirror size (4~m in comparison to 7~m), but also due to the wider separations between SSTs reducing the stereo trigger rate and event statistics. Nevertheless, a precision of $< 3\%$ is still achieved on these SSTs, with the MST subsystem calibrated to  $< 1\%$ and the LST subsystem precision level limited by a systematic offset from the scaling ratio. As expected, a higher level of accuracy is achieved with the more compact 2A array than for 2B, due to the larger event statistics occuring with lower telescope separations. With the amount of $\gamma$-ray showers used for this calibration, the remaining systematics in the method are already the dominating factor in the level of precision obtained. From figure \ref{fig:effagree}, it can clearly be seen that all telescopes are individually calibrated to $< 10\%$  with this method, the majority to $< 5\%$. This meets the CTA performance goal of $5\%$ systematic uncertainty on the individual telescope measurement of the absolute Cherenkov light intensity \cite{Gaug14}, whilst the SSTs may struggle to achieve this with alternatives such as the muon calibration.  

\noindent All results presented here are for $\gamma$--ray events only, whereas in practice IACT arrays suffer from a huge hadronic background. To mitigate this, $\gamma$--hadron separation procedures are routinely performed, such as the use of image shape cuts. Corresponding proton simulations were run using the same telescope configurations; assuming that the hadronic background follows the Cosmic Ray spectrum, these corresponded to an observing time of just a few seconds. 

Applying shape cuts based only on information from the two telescopes being compared in each pair measurement results in the selection of $\gamma$--like subshowers, in which case systematics from the poor performance of the energy reconstruction dominate. In a more realistic background rejection scenario, shape cuts are applied based on information from all triggered telescopes to select $\gamma$--like hadronic showers. This improved background rejection leaves just $\sim O( 10^{-3})\%$ of $7.95 \times 10^{9}$ simulated events passing cuts.
Despite this, encouraging results were obtained, with a precision of $11\%$ ($13\%$) obtained for the MST subsystem of layout 2A (2B), whilst for the LST (much better background rejection) and SST (much lower trigger rate) subsystems, meaningful conclusions could not be drawn from the surviving statistics.

\noindent Simulating sufficient hadronic background to achieve the same level of precision as that quoted for $\gamma$--rays in table \ref{tab:arrayrmss} is currently prohibitively expensive in computing time and memory. However, once data taking with the first CTA telescopes commences, the overwhelming hadronic background will provide ample opportunity for a thorough characterisation of this procedure prior to its final implementation on the full array.
The image size asymmetry procedure as outlined in \citep{Hofmann03}, may also be used as an additional cross check.

\section{Conclusion}
\label{sec:conclusion}

\noindent Cross calibration of telescope response efficiencies through the use of $\gamma$-ray images has been shown to be a robust approach enabling an independent calibration of many different hardware technologies. Relative calibration through pairwise comparisons ensures that multiple independent measurements overdetermine the system of unknown parameters, leading to an overall precision at the $1-3\%$ level after reasonable data collection times. This is within the CTA performance requirements on the telescope optical efficiency calibration for all telescope types. A slight bias was seen in energy asymmetry measurements involving telescopes with efficiencies quite far (up to 30\%) from the mean of the lookups. However, this was not seen to bias the response efficiencies recovered from this calibration prodecure significantly. This level of discrepancy between the true telescope response efficiency and the reference value in the lookup tables is unlikely to be encountered in practice, as regular monitoring should ensure that telescope response efficiencies remain fairly well known, with lookup tables being produced as required to more closely match the telescope parameters.

This method is capable of testing linearity under various intensities and intrinsically uses the correct input light spectrum, whereas the received light yield from muons is tied to that of a minimum ionising particle and has a slightly different Cherenkov light spectrum due to the difference in the production altitude of the Cherenkov light. Application of this method to data taken under various zenith angles and observing conditions would help eliminate possible angular or wavelength dependencies of the response coefficients. 

\noindent Further studies of the robustness of this approach could include various realistic scenarios of systematics (already the limiting factor in the level of precision obtained), such as a gradient in response efficiencies over an array, or of varying telescope optical throughput efficiency degradation rates. Situations involving biases due to the performance of the energy reconstruction can also be imagined, the influence of which merits further investigation. As the response efficiencies recovered via this procedure are relative, it is envisaged to be used alongside the other calibration methods.  Regardless of final implementation, this cross calibration procedure offers an ideal complement to the muon calibration for the next generation Cherenkov Telescope Array. In principle, such a method could be applied to other air shower physics experiments comprising multiple detector types situated at the same site.

\vspace{5pt}
\noindent This paper has undergone internal review by the CTA Consortium.
\vspace{10pt}

\noindent \textbf{Acknowledgements}
\newline
\noindent \begin{itshape}
We gratefully acknowledge financial support from:
The International Max Planck Research School for Astronomy and Cosmic Physics at the University of Heidelberg; and the Max Planck Society, Germany.
\end{itshape}

\bibliographystyle{ieeetr}
\bibliography{InterCalRefs.bib}

\begin{thebibliography}{10}

\bibitem{HintonHofmann09}
J.~A. {Hinton} and W.~{Hofmann}, ``{Teraelectronvolt Astronomy},'' {\em Annual
  Review of Astronomy \& Astrophysics}, vol.~47, pp.~523--565, Sept. 2009.

\bibitem{Hinton04}
J.~A. {Hinton} and {the HESS Collaboration}, ``{The status of the HESS
  project},'' {\em New Astronomy Reviews}, vol.~48, pp.~331--337, Apr. 2004.

\bibitem{Lorenz04}
E.~{Lorenz} and {The MAGIC Collaboration}, ``{Status of the 17 m {\empty} MAGIC
  telescope},'' {\em New Astronomy Reviews}, vol.~48, pp.~339--344, Apr. 2004.

\bibitem{Weekes02}
T.~C. {Weekes}, H.~{Badran}, S.~D. {Biller}, I.~{Bond}, S.~{Bradbury},
  J.~{Buckley}, D.~{Carter-Lewis}, M.~{Catanese}, S.~{Criswell}, W.~{Cui},
  P.~{Dowkontt}, C.~{Duke}, D.~J. {Fegan}, J.~{Finley}, L.~{Fortson},
  J.~{Gaidos}, G.~H. {Gillanders}, J.~{Grindlay}, T.~A. {Hall}, K.~{Harris},
  A.~M. {Hillas}, P.~{Kaaret}, M.~{Kertzman}, D.~{Kieda}, F.~{Krennrich}, M.~J.
  {Lang}, S.~{LeBohec}, R.~{Lessard}, J.~{Lloyd-Evans}, J.~{Knapp},
  B.~{McKernan}, J.~{McEnery}, P.~{Moriarty}, D.~{Muller}, P.~{Ogden},
  R.~{Ong}, D.~{Petry}, J.~{Quinn}, N.~W. {Reay}, P.~T. {Reynolds}, J.~{Rose},
  M.~{Salamon}, G.~{Sembroski}, R.~{Sidwell}, P.~{Slane}, N.~{Stanton}, S.~P.
  {Swordy}, V.~V. {Vassiliev}, and S.~P. {Wakely}, ``{VERITAS: the Very
  Energetic Radiation Imaging Telescope Array System},'' {\em Astroparticle
  Physics}, vol.~17, pp.~221--243, May 2002.

\bibitem{Vacanti94}
G.~{Vacanti}, P.~{Fleury}, Y.~{Jiang}, E.~{Par{\'e}}, A.~C. {Rovero},
  X.~{Sarazin}, M.~{Urban}, and T.~C. {Weekes}, ``{Muon ring images with an
  atmospheric {\v C}erenkov telescope},'' {\em Astroparticle Physics}, vol.~2,
  pp.~1--11, Feb. 1994.

\bibitem{Bolz04}
O.~Bolz, {\em {Absolute Energiekalibration der abbildenden Cherenkov-Teleskope
  des H.E.S.S. Experiments und Ergebnisse erster Beobachtungen des
  Supernova-\"{U}berrests RX J1713.7-3946}}.
\newblock PhD thesis, Ruprecht-Karls-Universit\"{a}t Heidelberg, 2004.

\bibitem{Acharya13}
B.~S. {Acharya}, M.~{Actis}, T.~{Aghajani}, G.~{Agnetta}, J.~{Aguilar},
  F.~{Aharonian}, M.~{Ajello}, A.~{Akhperjanian}, M.~{Alcubierre},
  J.~{Aleksi{\'c}}, {\em et~al.}, ``{Introducing the CTA concept},'' {\em
  Astroparticle Physics}, vol.~43, pp.~3--18, Mar. 2013.

\bibitem{DaviesCotton57}
J.~M. {Davies} and E.~S. {Cotton}, ``{Design of the quartermaster solar
  furnace},'' {\em Solar Energy}, vol.~1, pp.~16--22, Apr. 1957.

\bibitem{Vassiliev08}
V.~V. {Vassiliev} and S.~J. {Fegan}, ``{Schwarzschild-Couder two-mirror
  telescope for ground-based {$\gamma$}-ray astronomy},'' {\em International
  Cosmic Ray Conference}, vol.~3, pp.~1445--1448, 2008.

\bibitem{Hofmann03}
W.~{Hofmann}, ``{Intercalibration of Cherenkov telescopes in telescope
  arrays},'' {\em Astroparticle Physics}, vol.~20, pp.~1--3, Oct. 2003.

\bibitem{Gast09}
R.~Gast, ``{Investigating Systematics in the Energy Reconstruction of the
  H.E.S.S. telescopes},'' Master's thesis, Ruprecht-Karls-Universit\"{a}t
  Heidelberg, 2009.

\bibitem{Aleksic12}
J.~{Aleksi{\'c}} {\em et~al.}, ``{Performance of the MAGIC stereo system
  obtained with Crab Nebula data},'' {\em Astroparticle Physics}, vol.~35,
  pp.~435--448, Feb. 2012.

\bibitem{Gaug14}
M.~{Gaug}, D.~{Berge}, M.~{Daniel}, M.~{Doro}, A.~{F{\"o}rster}, W.~{Hofmann},
  M.~C. {Maccarone}, R.~D. {Parsons}, R.~{de los Reyes Lopez}, and C.~{van
  Eldik}, ``{Calibration strategies for the Cherenkov Telescope Array},'' in
  {\em Society of Photo-Optical Instrumentation Engineers (SPIE) Conference
  Series}, vol.~9149 of {\em Society of Photo-Optical Instrumentation Engineers
  (SPIE) Conference Series}, p.~19, Aug. 2014.

\bibitem{Hillas85}
A.~M. {Hillas}, ``{Cerenkov light images of EAS produced by primary gamma},''
  {\em International Cosmic Ray Conference}, vol.~3, pp.~445--448, Aug. 1985.

\bibitem{Lebohec98}
S.~{Le Bohec}, B.~{Degrange}, M.~{Punch}, A.~{Barrau}, R.~{Bazer-Bachi},
  H.~{Cabot}, L.~M. {Chounet}, G.~{Debiais}, J.~P. {Dezalay},
  A.~{Djannati-Atai}, D.~{Dumora}, P.~{Espigat}, B.~{Fabre}, P.~{Fleury},
  G.~{Fontaine}, R.~{George}, C.~{Ghesquiere}, P.~{Goret}, C.~{Gouiffes}, I.~A.
  {Grenier}, L.~{Iacoucci}, I.~{Malet}, C.~{Meynadier}, F.~{Munz}, T.~A.
  {Palfrey}, E.~{Pare}, Y.~{Pons}, J.~{Quebert}, K.~{Ragan}, C.~{Renault},
  M.~{Rivoal}, L.~{Rob}, P.~{Schovanek}, D.~{Smith}, J.~P. {Tavernet}, and
  J.~{Vrana}, ``{A new analysis method for very high definition imaging
  atmospheric Cherenkov telescopes as applied to the CAT telescope.},'' {\em
  Nuclear Instruments and Methods in Physics Research A}, vol.~416,
  pp.~425--437, Oct. 1998.

\bibitem{Bernlohr13ICRC}
K.~{Bernl{\"o}hr}, A.~{Barnacka}, Y.~{Becherini}, O.~{Blanch Bigas},
  A.~{Bouvier}, E.~{Carmona}, P.~{Colin}, G.~{Decerprit}, F.~{Di Pierro},
  F.~{Dubois}, C.~{Farnier}, S.~{Funk}, G.~{Hermann}, J.~A. {Hinton}, T.~B.
  {Humensky}, T.~{Jogler}, B.~{Kh{\'e}lifi}, T.~{Kihm}, N.~{Komin}, J.-P.
  {Lenain}, R.~{L{\'o}pez-Coto}, G.~{Maier}, D.~{Mazin}, M.~C. {Medina},
  A.~{Moralejo}, R.~{Moderski}, S.~J. {Nolan}, S.~{Ohm}, E.~{de O{\~n}a
  Wilhelmi}, R.~D. {Parsons}, M.~{Paz Arribas}, G.~{Pedaletti}, S.~{Pita},
  H.~{Prokoph}, C.~B. {Rulten}, U.~{Schwanke}, M.~{Shayduk}, V.~{Stamatescu},
  P.~{Vallania}, S.~{Vorobiov}, R.~{Wischnewski}, M.~{Wood}, T.~{Yoshikoshi},
  A.~{Zech}, and f.~t. {CTA Consortium}, ``{Progress in Monte Carlo design and
  optimization of the Cherenkov Telescope Array},'' in {\em {Proceedings of the
  33$^\mathrm{rd}$ International Cosmic Ray Conference}}, July 2013.

\bibitem{Bernlohr13}
K.~{Bernl{\"o}hr}, A.~{Barnacka}, Y.~{Becherini}, O.~{Blanch Bigas},
  E.~{Carmona}, P.~{Colin}, G.~{Decerprit}, F.~{Di Pierro}, F.~{Dubois},
  C.~{Farnier}, S.~{Funk}, G.~{Hermann}, J.~A. {Hinton}, T.~B. {Humensky},
  B.~{Kh{\'e}lifi}, T.~{Kihm}, N.~{Komin}, J.-P. {Lenain}, G.~{Maier},
  D.~{Mazin}, M.~C. {Medina}, A.~{Moralejo}, S.~J. {Nolan}, S.~{Ohm}, E.~{de
  O{\~n}a Wilhelmi}, R.~D. {Parsons}, M.~{Paz Arribas}, G.~{Pedaletti},
  S.~{Pita}, H.~{Prokoph}, C.~B. {Rulten}, U.~{Schwanke}, M.~{Shayduk},
  V.~{Stamatescu}, P.~{Vallania}, S.~{Vorobiov}, R.~{Wischnewski},
  T.~{Yoshikoshi}, A.~{Zech}, and {CTA Consortium}, ``{Monte Carlo design
  studies for the Cherenkov Telescope Array},'' {\em Astroparticle Physics},
  vol.~43, pp.~171--188, Mar. 2013.

\bibitem{James75}
F.~{James} and M.~{Roos}, ``{Minuit - a system for function minimization and
  analysis of the parameter errors and correlations},'' {\em Computer Physics
  Communications}, vol.~10, pp.~343--367, Dec. 1975.

\bibitem{Aharonian06}
F.~{Aharonian}, A.~G. {Akhperjanian}, A.~R. {Bazer-Bachi}, M.~{Beilicke},
  W.~{Benbow}, D.~{Berge}, K.~{Bernl{\"o}hr}, C.~{Boisson}, O.~{Bolz},
  V.~{Borrel}, I.~{Braun}, F.~{Breitling}, A.~M. {Brown}, R.~{B{\"u}hler},
  I.~{B{\"u}sching}, S.~{Carrigan}, P.~M. {Chadwick}, L.-M. {Chounet},
  R.~{Cornils}, L.~{Costamante}, B.~{Degrange}, H.~J. {Dickinson},
  A.~{Djannati-Ata{\"i}}, L.~{O'C.~Drury}, G.~{Dubus}, K.~{Egberts},
  D.~{Emmanoulopoulos}, P.~{Espigat}, F.~{Feinstein}, E.~{Ferrero},
  A.~{Fiasson}, G.~{Fontaine}, S.~{Funk}, S.~{Funk}, Y.~A. {Gallant},
  B.~{Giebels}, J.~F. {Glicenstein}, P.~{Goret}, C.~{Hadjichristidis},
  D.~{Hauser}, M.~{Hauser}, G.~{Heinzelmann}, G.~{Henri}, G.~{Hermann}, J.~A.
  {Hinton}, W.~{Hofmann}, M.~{Holleran}, D.~{Horns}, A.~{Jacholkowska}, O.~C.
  {de Jager}, B.~{Kh{\'e}lifi}, N.~{Komin}, A.~{Konopelko}, K.~{Kosack}, I.~J.
  {Latham}, R.~{Le Gallou}, A.~{Lemi{\`e}re}, M.~{Lemoine-Goumard}, T.~{Lohse},
  J.~M. {Martin}, O.~{Martineau-Huynh}, A.~{Marcowith}, C.~{Masterson},
  T.~J.~L. {McComb}, M.~{de Naurois}, D.~{Nedbal}, S.~J. {Nolan}, A.~{Noutsos},
  K.~J. {Orford}, J.~L. {Osborne}, M.~{Ouchrif}, M.~{Panter}, G.~{Pelletier},
  S.~{Pita}, G.~{P{\"u}hlhofer}, M.~{Punch}, B.~C. {Raubenheimer}, M.~{Raue},
  S.~M. {Rayner}, A.~{Reimer}, O.~{Reimer}, J.~{Ripken}, L.~{Rob},
  L.~{Rolland}, G.~{Rowell}, V.~{Sahakian}, L.~{Saug{\'e}}, S.~{Schlenker},
  R.~{Schlickeiser}, U.~{Schwanke}, H.~{Sol}, D.~{Spangler}, F.~{Spanier},
  R.~{Steenkamp}, C.~{Stegmann}, G.~{Superina}, J.-P. {Tavernet}, R.~{Terrier},
  C.~G. {Th{\'e}oret}, M.~{Tluczykont}, C.~{van Eldik}, G.~{Vasileiadis},
  C.~{Venter}, P.~{Vincent}, H.~J. {V{\"o}lk}, S.~J. {Wagner}, and M.~{Ward},
  ``{Observations of the Crab nebula with HESS},'' {\em Astronomy and
  Astrophysics}, vol.~457, pp.~899--915, Oct. 2006.

\end{thebibliography}

\end{document}